\documentclass[%
 reprint,
superscriptaddress,
 amsmath,amssymb,
 aps,
 prl,
]{revtex4-2}

\usepackage{graphicx}
\usepackage{dcolumn}
\usepackage{bm}

\usepackage{balance}

\usepackage{epstopdf}

\begin{document}

\title{How a leak can stop itself}

\title[How a leak can stop itself]{How a leak can stop itself}

\author{Caroline D. Tally}
\altaffiliation{These authors contributed equally to this work.}
\affiliation{
Department of Physics, Williams College, Williamstown, MA 01267, USA
}

\author{Heather E. Kurtz}
\altaffiliation{These authors contributed equally to this work.}
\affiliation{
Department of Physics, Williams College, Williamstown, MA 01267, USA
}

\author{Rose B. Tchuenkam}
\affiliation{
Department of Physics, Williams College, Williamstown, MA 01267, USA
}

\author{Justyn M. Friedler}
\affiliation{
Department of Physics, Williams College, Williamstown, MA 01267, USA
}

\author{Katharine E. Jensen}
\email{kej2@williams.edu} 
\affiliation{
Department of Physics, Williams College, Williamstown, MA 01267, USA
}

\date{\today}

\begin{abstract}
Small fluid leaks are common and frequently troublesome. 
We often consider how to stop a leak, but here we ask a different question: 
how might a leak stop itself? 
We experimentally study leaking flow transitions from continuous drainage to spontaneous arrest. 
High-speed imaging reveals that fluid breakup events generate droplets whose Laplace pressures oppose the leak. 
Early droplets grow unstably, allowing the leak to continue, but ultimately a final capping droplet equilibrates to a stable spherical cap via lightly damped harmonic oscillations. 
A total energetic theory incorporating both the potential and kinetic energy of attempted capping droplets shows that inertia plays a key role in the leak-stop mechanism.
Experiments examining the stability of rivulet flow in such a system demonstrate that a transition from continuous to discrete flow is an essential prerequisite in determining when a leak can stop itself.
\end{abstract}

\maketitle

\section{Introduction}

When a tube springs a leak, the motion of the exiting liquid depends on fluid and vessel material properties, geometry and orientation of the hole, and driving pressure \cite{extrand2018liquid,extrand2018drainage}.
Flow transitions in the related process of pouring from a container, and in particular the ``teapot effect'', have inspired decades of physics research \cite{reiner1956teapot,keller1957teapot,reba1966applications,walker1984troublesome,duez2010wetting,dong2015manipulating,dong2015superwettability,ferrand2016wetting,shi2018fluid,jambon2019liquid}. 
Meanwhile, our understanding of the fundamental physical processes that govern transitions between different regimes of a leaking flow---defined as a low-volume fluid flow exiting from a constrained geometry---remains incomplete.

Of particular interest is the possibility that an active leak could spontaneously arrest even against a nonzero driving pressure.
Substantial work has focused on developing methodologies for detecting and evaluating fluid leaks, especially in large-scale systems \cite{bialous1969leakage,jackson1998leaktesting,colombo2009selective,zhang2013review}, 
and the study of leak-like fluid flows in the contexts of imbibition and drainage in porous media is an area of active research 
\cite{washburn1921dynamics, perkins1963review, datta2013drainage,huynh2017millimeter,extrand2018comment, lu2021forced, esser2021network}. 
Understanding the governing physics that determines the moment and mechanism by which a leak might stop itself spontaneously could lead to novel pipe fitting designs \cite{extrand2018drainage2,extrand2018drainage3}, engineering failsafes for fluidic systems, and new fundamental insights into fluid flow from constrained geometries.
It is well-established that small holes in a material can effectively be ``capped'' by the fluid-air interfaces themselves, whose surface tension, curvature, and resulting Laplace pressure\cite{deGennes2004} oppose leakage; 
this forms the physical basis for engineering waterproof yet breathable fabrics like Gore-Tex.\cite{gohlke1976gore}
Recent experiments measured the hydrostatic driving pressures for leaking fluid from a small hole and noted that the leaks ultimately stopped with a spherical cap of fluid spanning the hole, but observed leaks persisting much longer than predicted and did not examine the mechanism of spontaneous arrest.\cite{extrand2018drainage}
The fundamental mechanics of why, how, and when such a capping droplet could emerge from a dynamic flow has not yet been investigated.

In this paper, we use high-speed imaging to characterize a small fluid leak draining from a circular hole in the side of a vertically-oriented tube.
We observe the leak as it progresses through a series of flow transitions from continuous drainage to spontaneous arrest.
Fluid breakoff events during a discrete dripping regime generate small droplets of fluid whose Laplace pressure opposes further leakage.
Despite this, early droplets continue to grow unbounded, allowing the leak to continue for some time.
Ultimately, a final droplet equilibrates via lightly damped harmonic volume oscillations to form a stable spherical cap of fluid that stops the leak.
We develop stability and energetic analyses that take into account leak geometry, fluid and interface material properties, and the initial volume and growth rate of the droplets, and find that the initial kinetic energy of newly-formed droplets plays a key role in determining which droplet is able to stop the leak.
Finally, we note that the transition from continuous rivulet flow to discrete dripping is a key step in the progression to spontaneous self-arrest, and report first experiments examining how rivulet stability in these systems affects when a leak can stop itself. 

\section{Leaking Experiment}
We investigate leaking flow transitions using a simple experimental setup consisting of a 15-mL centrifugation tube (inner radius $r_c = 7.60$ mm; wall thickness 1.62 mm) 
with a small hole drilled in its side. 
For the experiments described here, we use pure, deionized (DI) water as the leaking fluid.
We modify the wettability of the tube either by changing the tube material from polypropylene (DI water contact angle $\theta = 102.1^\circ$) to polystyrene ($\theta = 87.4^\circ$), or 
by coating the exterior surface of a polypropylene tube with solid paraffin (Parafilm-M) ($\theta = 108.9^\circ$) or NeverWet superhydrophobic treatment \cite{neverwet2015} ($\theta = 149^\circ$).
We use moderately to highly hydrophobic exterior surfaces ($\theta \gtrsim 90^\circ$) in order to avoid complications from a sliding-droplet leaking mechanism described in earlier literature.\cite{extrand2018drainage} 
All contact angles are measured using the sessile drop method.\cite{deGennes2004,yarnold1949angle}
The fluid viscosity can also be increased by using water-glycerol solutions. 
We drill a hole in each tube using an end mill of the desired size and carefully de-burr to minimize roughness.
The hole radius $r_h$ is varied between 0.4 mm and 1.6 mm, always less than the capillary length of water ($l_c = 2.7$ mm).
The tube is cleaned with isopropyl alcohol and soaked in DI water to remove any possible contamination that could affect fluid surface energies, 
then mounted and aligned vertically with respect to gravity using a plumb line.

We initiate a leak by filling the tube with DI water well above the hole so that the filling conditions do not affect subsequent flow behavior.
The weight of the fluid above the hole provides a hydrostatic driving pressure $P_\text{drive}$.
For draining experiments, the tube is open at the top and the small hole, and the slow decrease of the hydrostatic pressure head during leaking provides a smoothly decreasing $P_\text{drive}$ over the course of an experiment.
We include overview images of an example draining experiment in supplementary Fig. S1.$^\dag$
We also constructed a modified leaking setup that allows us to add water to the tubes from the bottom, far from the small hole, at specified filling rates using a syringe pump (Harvard Apparatus Pump 11 Elite).
This filling setup enables us to examine both the onset and cessation of leaking flows at controlled average flow rates, as discussed below.

The leaks are imaged at slower frame rates (41 fps) and larger field of view using a camera (Thorlabs DCC3260C) with macro lens and uniform back-illuminating LED light source (Edmund Optics 83-873) in order to capture their full flow evolution from leak initiation to spontaneous arrest while simultaneously measuring the fluid level and hence hydrostatic pressure over time.
We use this setup to explore how leaks spontaneously arrest over a broader parameter space of fluid viscosity, wetting angle, and hole size. 
We use a high-speed camera (Phantom v310 or Photron Nova S16) equipped with a telecentric lens (Edmund Optics 88-386) to study specific flow transitions with high spatial and time resolution in a smaller number of complementary experiments.

In a typical draining experiment, we observe a progression through four distinct flow regimes:
(1) jetting, in which the fluid spouts parabolically away from the tube according to Torricelli's law; \cite{white2021fluidmechanics} 
(2) rivulet flow, in which a thin, continuous stream of fluid partially wets the tube exterior until eventually breaking up via a Rayleigh-Plateau-like instability;\cite{deGennes2004,rayleigh1879stability, davis1980moving,diez2009breakup} 
(3) discrete dripping; and, finally, (4) spontaneous arrest of the leak, with the final fluid level still above the hole.\cite{extrand2018drainage}
All leaks exhibited a jetting regime at high driving pressures, 
but whether and at what driving pressures a leak progressed through the subsequent flow regimes depended on the fluid viscosity, hole radius, and exterior surface wettability.
We observed that increasing the fluid viscosity from about 1 cP (pure water at room temperature) to as much as 10 cP by using glycerol-water solutions resulted in earlier transitions out of the jetting regime, as higher viscosities result in slower flow velocities for the same driving pressure, making it more difficult for the jetting fluid to arc cleanly away from the tube.
However, interestingly, we observed that changing the fluid viscosity had no measurable effect on the pressure at which the flow-stop transition occurs, suggesting that the transition to spontaneous arrest is not sensitive to the flow rate of the fluid out of the hole.\cite{CarolineTallyThesis2022}

By contrast, both hole radius $r_h$ and exterior contact angle $\theta$ strongly affect how and when a leak can stop itself. 
At small $r_h$ and high contact angle $\theta$, we observed that leaks were unable to establish a rivulet, and instead transitioned directly from jetting to extended dripping en route to spontaneous arrest.
Meanwhile, at lower $\theta$ and/or larger hole sizes, we observed leaks stopping directly from rivulet flow without going through an extended dripping phase.
(See supplementary Figs. S3 and S4$^\dag$ for a summary of leak-stop sequences observed across the parameter space of varied contact angle and hole size.)
In between, we observe the most general behavior of a spontaneously-arresting leak that goes through all four phases of flow behavior. 

\begin{figure}[htb]
    \centering
    \includegraphics[width=0.48\textwidth]{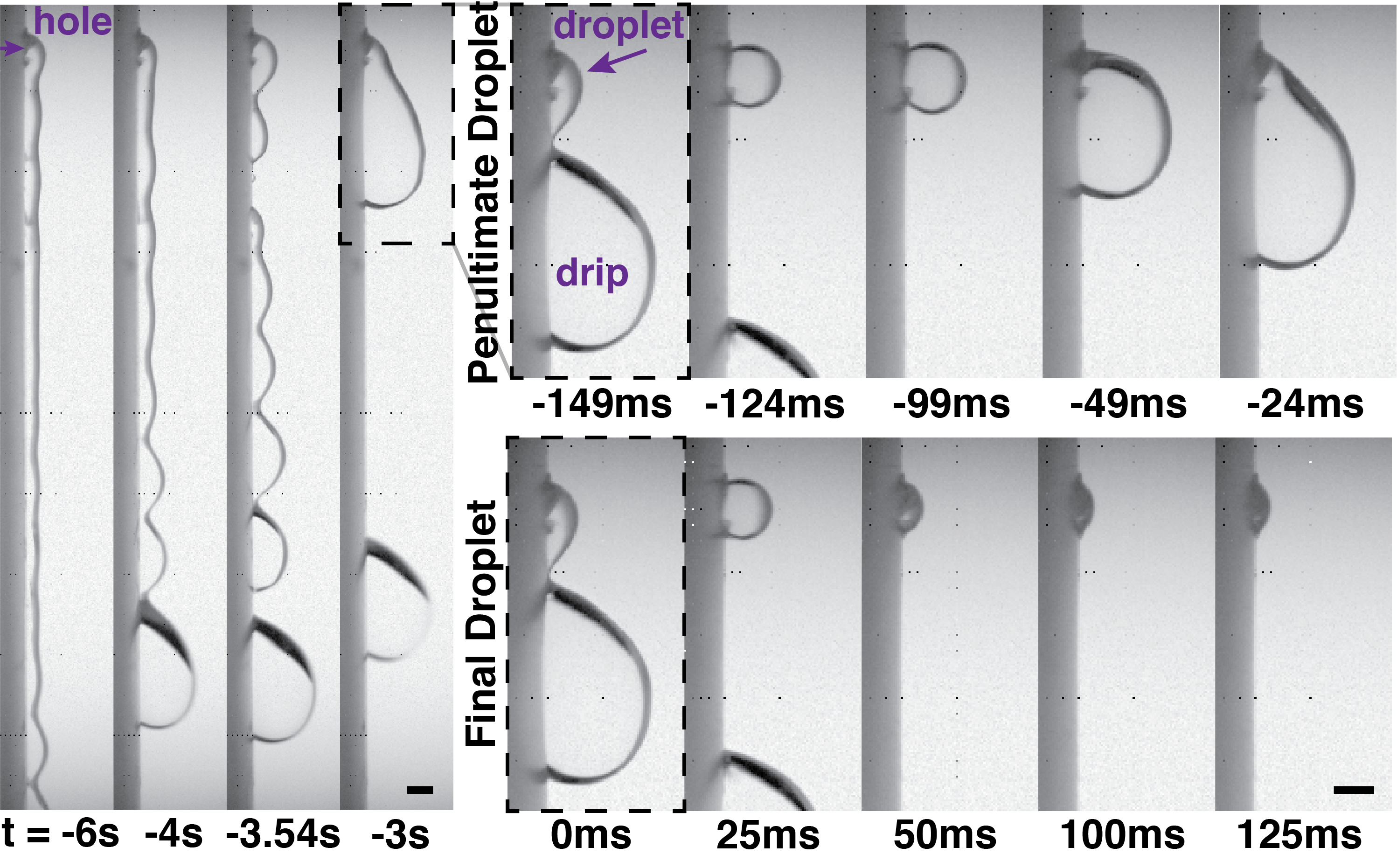}
    \caption{
    The evolution of a leaking flow from a small hole (indicated) in the side of a vertical tube, captured at 1000 fps.
    An initial jetting regime prior to establishing rivulet flow is not shown; 
    the leak then transitions from continuous rivulet flow through dripping to spontaneous arrest.
    After rivulet breakup, \cite{diez2009breakup}
    each drip breakoff generates a new droplet covering the hole whose Laplace pressure opposes further flow.
    Evolution of the final two droplets with frames aligned by time elapsed since the last breakoff event are shown at right.
    See also Supplementary Movies 1 and 1B.$^\dag$ 
    Time measured relative to the breakoff event that generates the final capping droplet.
    Scale bars equal 0.8 mm, the hole diameter.
    }
     \label{fig:stopdrop}
\end{figure}

We show example images from the final several seconds of such an experiment in Fig. \ref{fig:stopdrop}(b), including late-stage rivulet flow and breakup (left) through spontaneous arrest (right).
Time $t$ is measured relative to the breakoff event that produced the final capping droplet that stopped the leak.
In this experiment, the tube was coated with paraffin, chosen for its moderate hydrophobicity ($\theta > 90^\circ$),\cite{yarnold1949angle} and the hole radius was $r_h = 0.4$ mm.
This combination of parameters reliably produced an extended dripping regime prior to spontaneous arrest in repeated experiments.
These example frames were taken from a high-speed video acquired at 1000 frames per second with a 30 $\mu$s exposure time using a Phantom v310 high-speed camera. 
Raw videos are included in the Supplementary Information.$^\dag$ 
We focus the following analysis and discussion on this experimental setup that demonstrates all four flow regimes in order to elucidate the physical mechanisms that underpin how a leak can stop itself.

\section{Mapping Droplets from Dripping to Leak-Stop} 

Following rivulet breakup, each large drip that breaks off and falls away leaves behind a small droplet across the hole, as seen in Fig. \ref{fig:stopdrop}(b)(right).
The surface tension and curvature of the droplet result in a Laplace pressure that opposes the driving pressure.\cite{deGennes2004,extrand2018drainage} 
For leaks that exhibit an extended dripping regime, the back-pressure from early attempted capping droplets is initially insufficient to stop the leak;
rather, each droplet grows significantly over time until it is large enough to drip off, thus continuing the leak as well as generating the next attempted capping droplet. 
Eventually, however, we observe that a final droplet arrests the flow at a final equilibrium driving pressure $P_\text{equil}$ that is consistent across repeat experiments, in agreement with an earlier study that measured initiation and equilibrium driving pressures for leaks with a similar geometry.\cite{extrand2018drainage}

We compare the evolution of the final two attempted capping droplets on the right side of Fig. \ref{fig:stopdrop}(b) as well as in Supplementary Movie 1B,$^\dag$
with images aligned according to time since the previous breakoff.
The penultimate droplet, like all droplets before it, grows until breaking off as a much larger-volume drip, while the final droplet ultimately shrinks and equilibrates to stop the leak.
Difference images comparing equivalent times after each breakoff event show that the droplets decrease slightly in size with successive breakoff events, while the drips that break away are indistinguishable from each other for at least the final 20 breakoffs, similar to free dripping. \cite{tate1864xxx,tsai2019classification} 
The observed incremental reduction in droplet volume corresponds to an incremental decrease in fluid flow rate resulting from a decrease $\Delta P$ in the hydrostatic driving pressure $P_\text{drive}$ as each larger drip falls away.
By catching and weighing known numbers of drips, we measure the average drip volume to be $V_\text{drip} = 26 \pm 3$ $\mu$L (mean $\pm$ standard deviation), equivalent to a $\Delta P = 0.35 \pm 0.04$ Pa per drip that falls away.

In order to understand what enables the final droplet---but not its predecessors---to stop the leak, we measure the geometry of each droplet by mapping its surface as well as that of the flat exterior tube surface over time using edge-mapping software developed in our earier work.\cite{berman2019singular,jensen2015wetting} 
Newly-formed droplets recoil from breakoff and quickly assume a spherical cap geometry,\cite{weisstein2008spherical} 
shown schematically in supplementary Fig. S2.$^\dag$
Even ultimately-unstable droplets grow as axisymmetric spherical caps for some time after breakoff before becoming large enough that gravity breaks this symmetry.

From the mapped surface profiles over time, we directly measure the radius of curvature $R$ and position of each axisymmetric droplet relative to the flat exterior tube surface by fitting a circle to the central part of the droplet profiles.
From these data, we then determine the contact radius $a$, distance from the flat tube surface to the peak of the spherical cap $d$, and droplet volume $V$. 
Since we measure individual droplets only while they remain axisymmetric (up to a few $\mu$L in volume, much smaller than the drips that fall away) the change in hydrostatic $P_\text{drive}$ during measured droplet growth is negligible ($\Delta P<0.05$ Pa).

\begin{figure}[t!]
    \centering
    \includegraphics[width=0.47\textwidth]{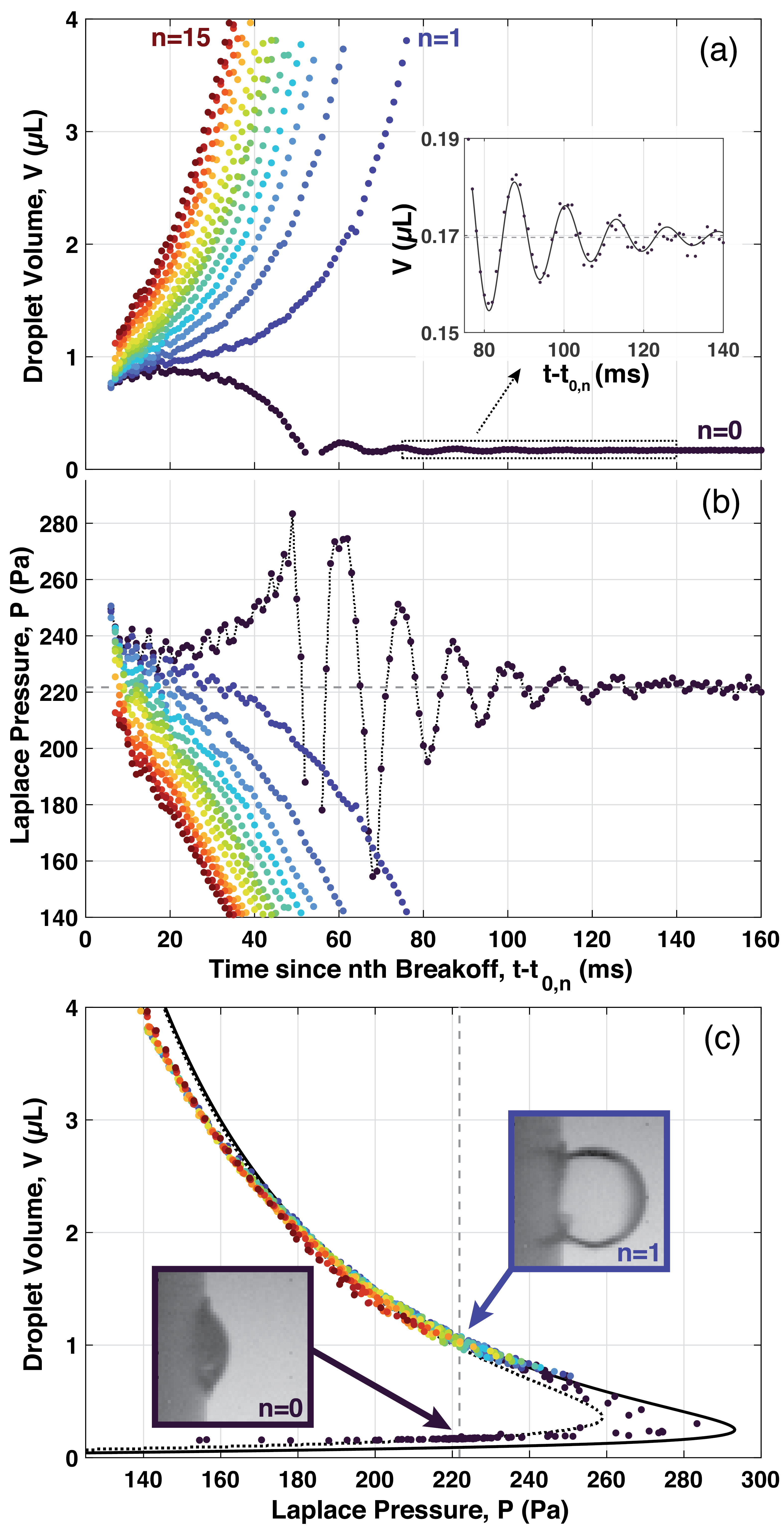}
    \caption{
    Droplet volume $V$ and Laplace pressure $P$ for the final 16 droplets of the leaking experiment shown in Fig. \ref{fig:stopdrop}. 
    Data colored dark blue to dark red by droplet number. 
    (a) $V$ vs. time since previous drip breakoff, $(t-t_{0,n})$.
    Inset: $n=0$ data during equilibration, with exponentially decaying cosine fit (solid line) and $V_\text{equil} = 0.17$ $\mu$L indicated (dashed).
    (b) $P$ vs. $(t-t_{0,n})$ for the same droplets, with $P_\text{equil} = 222$ Pa indicated.
    (c) $V$ vs. $P$, with theoretical curves plotted for $a_{i}$ (solid) and $a_\text{equil}$ (dotted).
    Example images of the final ($n=0$, stable) and penultimate ($n=1$, unstable) droplets each with $P = P_\text{drive}$ are shown inset.
    }
     \label{fig:PV}
\end{figure}

We plot the droplet volume $V$ vs. time since the previous breakoff event, $t-t_{0,n}$, in Fig. \ref{fig:PV}(a) for the final 16 droplets of the experiment shown in Fig. \ref{fig:stopdrop}, 
colored according to droplet number $n$ from the final capping droplet ($n=0$) in dark blue up to an earlier droplet $n=15$ in red.
All droplets grow immediately after formation because they are dynamically created while fluid continues to flow out of the hole.
As the driving pressure decreases incrementally with decreasing droplet number $n$, the flow rate also slows, which is reflected both in the continuous decrease in initial droplet volume as well as the decrease in initial volumetric growth rate of successive droplets.
After formation, each of the last $\sim 10$ droplets experiences an initial slowdown in its growth rate, but for all but the last one ($n \geq 1$), the growth subsequently accelerates when the droplet volume exceeds about 1 $\mu$L.
The final droplet, by contrast, reaches a maximum volume of 0.89 $\mu$L, then shrinks at an accelerating rate, significantly undershooting its equilibrium volume and even briefly disappearing fully back into the hole between 53-55 ms.
After reemerging, the final droplet equilibrates via damped harmonic volume oscillations to a final, stable capping droplet volume, as shown in Fig. \ref{fig:PV}(a,Inset).

The droplet Laplace pressures $P = 2 \gamma/R$ over time provide insight into the force balance that drives the time evolution of droplet volumes.
We calculate $P$ using the surface tension of pure water, $\gamma = 72$ mN/m, and the measured capping droplet radii $R$, and plot $P$ vs. $t-t_{0,n}$ in Fig. \ref{fig:PV}(b).
The maximum possible Laplace pressure for a capping droplet occurs at the smallest possible radius of curvature, $R=a$, where the droplet is a hemisphere;
this also establishes a critical driving pressure $P_{crit} = 2 \gamma/a$ above which it is impossible for a capping droplet to block the leak.
We confirmed using tube-filling experiments with slowly increasing pressure that a tube with a stable capping droplet begins to leak when $P_{drive} \geq P_{crit}$, in agreement with previous literature,\cite{extrand2018drainage} with leak initiation starting via a sudden ``dam-bursting'' from the capping droplet. 

In Fig. \ref{fig:PV}(b) we see that the final droplet equilibrates to a Laplace pressure $P_\text{equil} = 222$ Pa. 
This corresponds to balancing a hydrostatic driving pressure with fluid height $H = P_\text{equil}/(\rho g) = 22.7$ mm above the hole, where $\rho = 997$ kg/m$^3$ is the density of water at room temperature and $g$ is the gravitational constant.
While this remains a significant driving pressure, it is substantially below the initial critical pressure $P_{crit} = 2 \gamma/a_i = 294$ Pa when the droplet was formed. 
The dam-bursting flow initiation mechanism is very different than the equilibrating capping droplet mechanism we observe during spontaneous arrest, so it is reasonable that $P_\text{equil}$ might not be the same as $P_{crit}$. 
However, we see from Fig. \ref{fig:PV}(b) that several droplets prior to $n=0$ started with $P > P_\text{drive}$, but were nonetheless unable to stop the leak, even accounting for the slightly higher driving pressures $P_\text{drive} = P_\text{equil} + n \Delta P$ experienced by earlier droplets.
Why did the final capping droplet successfully stop the leak when the previous ones failed to do so?

It is useful to note that two capping droplets can have the same Laplace pressure while having very different volumes, as shown schematically in supplementary Fig. S2(b-c);$^\dag$
such a pair of complementary spherical caps can be obtained by slicing a sphere of radius $R$ using a plane that intersects the original sphere over an area $\pi a^2$.
We plot $V$ vs. $P$ for the same 16 droplets in Fig. \ref{fig:PV}(c), 
and indeed observe distinct upper (larger volume, $V_L$) and lower (smaller volume, $V_s$) branches that meet at a maximum pressure.
Geometrically, all other droplets with contact radius $a$ must have a larger radius of curvature $R>a$, and hence a smaller Laplace pressure, but can have a larger or smaller volume than the hemisphere. 
Interestingly, we measure all droplets to have the same initial contact radius $a_i = 0.491 \pm 0.005$ mm when they are first created, about 20\% larger than the hole radius.
The final capping droplet wets out further as it goes through dramatic volume changes en route to stabilizing, then equilibrates with a new constant contact radius $a_\text{equil} = 0.556 \pm 0.002$ mm. 
Noting that $V_s + V_L = \frac{4}{3}\pi R^3$, we derive analytic expressions for $V_L(P,a)$ and $V_s(P,a)$ (detailed in the Supplementary Information$^\dag$) and plot these curves for contact radii $a=a_i$ (solid) and $a=a_\text{equil}$ (dotted) in Fig. \ref{fig:PV}(c).
We see good agreement between our data and analytic calculations, with deviations as expected both at large droplet volumes $V \gtrsim 2$ $\mu$L, where unstably growing droplets begin to wet out beyond $a_i$, and during the transition from $a_i$ to $a_\text{equil}$ for the final droplet.

Droplets on the upper, large-volume $V_L$ branch are always unstable.
Even when the Laplace pressure perfectly balances the driving pressure, $P = P_\text{drive}$, any perturbation that increases the droplet volume also decreases the droplet's Laplace pressure, leading to runaway growth.
The opposite is true on the $V_L$ branch for a perturbation that increases the droplet pressure by decreasing its volume;
this instead leads to runaway shrinkage of the droplet until it passes through the maximum-pressure point and crosses over to the lower branch.
On the lower, small-volume $V_s$ branch, by contrast, this feedback between $P$ and $V$ is reversed and $P = P_\text{drive}$ is a stable, attractive equilibrium.
We show snapshots of example droplets at $P = P_\text{drive}$ on both branches inset in Fig. \ref{fig:PV}(c). 
In our experiment, all droplets start on the upper, $V_L$ branch with a volume larger than the maximum-pressure hemisphere with $a=a_i$.
The final droplet is the only one that crosses over from the unstable to the stable branch of $V$ vs. $P$ and is thus able to stop the leak.

\section{Potential Energy Landscape of a Capping Droplet}

The time evolution of all of the droplets leading to spontaneous arrest suggests a potential energy landscape with an escapable basin of stability that is harmonic for small perturbations about the equilibrium volume.
Further, many droplets that begin with $P>P_\text{drive}$ nevertheless overshoot their equilibrium volume and fail to stop the leak, suggesting that inertia may play an important role in droplet growth and equilibration.
In order to understand our observations quantitatively, we consider the total energy of a control volume defined as all of the fluid above and in the hole, including the droplet.
We divide the total energy into an effective potential energy plus kinetic energy as $E_\text{tot} = U + K$.

\begin{figure}[htb]
    \centering
    \includegraphics[width=0.46\textwidth]{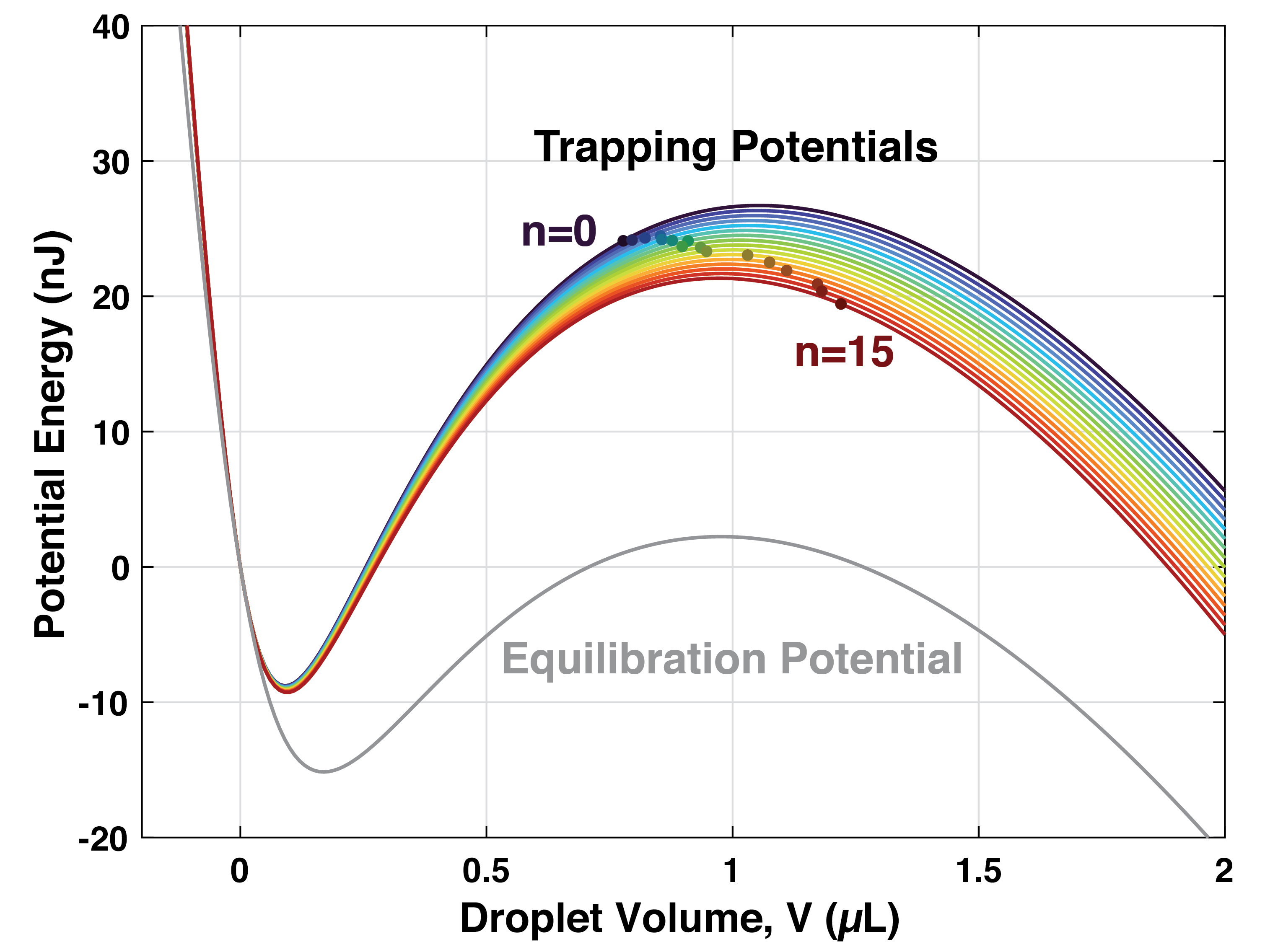}
    \caption{
    Droplet potential energy landscape.
    Colored lines show initial trapping potentials $U(V,a_i,P_{drive})$ from Eq. \ref{eqn:energy} for each of the last 16 droplets, colored by $n$ as above, with the driving pressure for each estimated as $P_{drive} = P_{equil} + n \Delta P$.
    Superimposed points are Eq. \ref{eqn:energy} evaluated for each droplet's measured volume at $t-t_{0,n} = 8$ ms.
    The potential in which the final droplet ($n=0$) equilibrated to the minimum energy, $U(V,a_{equil},P_{equil}$), is plotted as a gray line.
    }
    \label{fig:energy}
\end{figure}

We calculate the potential energy $U = U(V,a,P_\text{drive})$ accounting for both gravitational and surface energy contributions of the fluid in the control volume.
We take the zero energy configuration to be at droplet volume $V=0$, corresponding to a flat interface covering the hole with surface area $\pi a^2$, and a fluid height $H = P_\text{drive}/(\rho g)$.
As the droplet increases in volume, the height of the water inside the tube decreases while the surface area of the capping droplet increases by an amount $\pi d^2$, where $d = d(V,a)$ is the distance from the flat tube surface to the peak of the spherical cap, as detailed in the Supplementary Information (Eq. S6).$^\dag$ 
The resulting potential energy of a droplet is given by:
\begin{equation}
    U(V,a,P_\text{drive}) = -P_\text{drive} V 
    + \gamma \pi d^2 + \frac{\rho g }{2\pi r_c^2} V^2 
    \label{eqn:energy}
\end{equation}
\noindent The final term results from the small loss of hydrostatic pressure while filling the droplet.
For the droplet volumes we typically observe, less than a few $\mu$L, this term is extremely small ($\mathcal{O} \left(10^{-4} \right)$ compared to the others), 
but for completeness we retain it in all calculations. 
Equation \ref{eqn:energy} equivalently represents the work required to push out a droplet of volume $V$ under these conditions without changing the velocity field in the fluid, where $V$ is the droplet volume regardless of whether it is on the $V=V_L$ unstable branch or the $V=V_s$ stable branch.

Each droplet initially experiences a trapping potential defined by evaluating Eq. \ref{eqn:energy} for the initial contact radius $a_i$ and a unique driving pressure which we approximate as $P_\text{drive} = P_\text{equil} + n \Delta P$ to account for the incremental change in $P_\text{drive}$ with each successive drip.
We plot the initial trapping potentials $U(V,a_i,P_\text{drive})$ vs. $V$  for the last 16 droplets of this experiment in Fig. \ref{fig:energy}, again numbered with the final capping droplet being $n=0$ and colored by droplet number as above.
We also plot the equilibration potential $U(V,a_\text{equil},P_\text{equil})$ in which the final droplet equilibrates (gray line);
the larger contact radius $a_\text{equil}$ results in a shallower potential energy basin compared to the trapping potentials.
When the hydrostatic final term in Eq. \ref{eqn:energy} is negligible, the local extrema in the potential energy exactly correspond to pairs of capping droplets with the same Laplace pressure $P=P_\text{drive}$ but different volumes as described in Fig. \ref{fig:PV}(c): 
$V_s(P_\text{drive},a)$ at the stable minimum and  $V_L(P_\text{drive},a)$ at the unstable local maximum corresponding to the peak energy $U_\text{barrier}$.

Based on this energy landscape, we expect that any attempted capping droplet that starts with a volume $V_0 > V_L(P_\text{drive},a)$ will grow without bound and therefore continue the leak to the next breakoff event.
Droplets that start with a smaller $V_0 < V_L(P_\text{drive},a)$ can, in principle, be trapped into a stable, leak-stopping geometry, but only if they are not growing so quickly at the beginning that they escape over $U_\text{barrier}$.
In order to see where the observed attempted capping droplets start in this landscape, we plot as points the potential energy computed based on each droplet's measured volume at $t-t_{0,n} = 8$ ms, the earliest time at which every droplet was simultaneously well-fit by a spherical cap, allowing consistent volume comparisons.
The visible separation in volume at the measurement time $t-t_{0,n} = 8$ ms between the smallest 10 and the largest 6 droplets results because the droplet volumetric growth rate initially slowed at early times for the smallest droplets but always accelerated for the largest droplets, suggesting that these groups started on different sides of the potential energy barrier.

\section{Total Energy of Self-Arresting Leaks}

In agreement with droplet Laplace pressure measurements above (Fig. \ref{fig:PV}(b), several of the last attempted capping droplets were initially small enough to start within their respective trapping potential basins, but nonetheless escaped, suggesting that their total energy $E_\text{tot}$ exceeded $U_\text{barrier}$.
Thus far, we have only considered the droplets' potential energy $U(V,a_i,P_\text{drive})$;
however, because all droplets are formed during an active leak, they all have an initial, positive growth rate during approximately the first 10-20 ms after the fluid breakoff that created them, as shown in Fig. \ref{fig:PV}(a).
From an energetic perspective, this suggests that we should additionally consider the droplets' kinetic energy contribution to $E_\text{tot}$, which depends on the initial fluid flow rate. 

In order to evaluate the kinetic energy contribution to $E_\text{tot}$ rigorously, we would ideally integrate the squared fluid velocity over the entire control volume, but we lack a direct measurement of this velocity field.
However, we can take advantage of the observed harmonic volume oscillations of the final droplet to estimate the total kinetic energy by treating the system as a classical mass-and-spring damped harmonic oscillator with effective mass $m_\text{eff}$, effective spring constant $k_\text{eff}$, and natural frequency $\omega_0 = \sqrt{k_\text{eff}/m_\text{eff}}$, and initial kinetic energy 
\begin{equation}
   K = \frac{1}{2} m_{\text{eff}}  \left( \frac{dV}{dt} \right)^2
\end{equation}
\noindent where the volumetric growth rate $dV/dt$ is evaluated as close to droplet creation as possible.

In order to measure the effective mass $m_\text{eff}$ of the capping droplet, we first numerically evaluate $k_\text{eff} = \partial^2 U/\partial V^2|_{V_\text{equil}}$ 
for the equilibration potential at its minimum, corresponding to the equilibrium droplet volume, 
and obtain $k_\text{eff} = 570$ nJ/$\mu$L$^2$ for the potential shown in Fig. \ref{fig:energy}.
We next fit the volume oscillations of the final capping droplet for small displacements ($\pm$$<$$10\%$) about $V_\text{equil}$ with the decaying cosine function $V(t) = V_\text{equil} + A \cos{(\omega t + \phi)} e^{- t/\tau} $, as shown in Fig. \ref{fig:PV}(a, Inset).
This fit yields a measured oscillation frequency $\omega = 0.49$ ms$^{-1}$ and damping constant $1/\tau = 0.044$ ms$^{-1}$.
From these, we recover the natural frequency of the droplet volume oscillator as $\omega_0 \approx \omega$ because the damping is very light. 
Combining these results, we find the effective mass of this oscillator to be $m_\text{eff} = k_\text{eff}/\omega_0^2 = 2.3$ mg$/$mm$^4$.
The dimensions of $m_\text{eff}$ can be interpreted as [Density/Length],  
suggesting that a relevant length scale in determining the effective mass may be $l = \rho/m_{\text{eff}} = 0.42$ mm,  approximately equal to the hole radius.

Putting this all together, we estimate the total energy for a droplet with volume $V=V(t)$ to be 
\begin{equation}
    E_\text{tot} = U(V,a,P_\text{drive}) + \frac{1}{2} m_{\text{eff}} \left( \frac{dV}{dt} \right)^2
\end{equation}
\noindent and evaluate its 
excess energy above its trapping potential energy barrier as $E_\text{ex} = E_\text{tot} - U_\text{barrier}$, 
where $U_\text{barrier} = U(V_L(P_\text{drive},a),a,P_\text{drive})$ evolves for each droplet as the driving pressure incrementally decreases. 
We compute each droplet's early-time potential energy curve as in Fig. \ref{fig:energy} and fit a straight line to $V$ vs. $t-t_{0,n}$ (Fig. \ref{fig:PV}(a)) at early times up to $t-t_{0,n} = 12$ ms to approximate the early-time volumetric growth rates $dV/dt$.
These measured growth rates range from 0.014 $\mu$L/ms for droplet $n=0$ to 0.083 $\mu$L/ms for droplet $n=15$.
In order to quantify experiment-to-experiment variation, we additionally computed the potential, kinetic, and excess energy for an independent repeat of this experiment with the same hole size and exterior wettability;
measured droplet volumes and volumetric growth rates for the final 16 droplets from both experiments are included in Table S1$^\dag$, extracted geometric and oscillator parameters are included in Table S2$^\dag$, and both raw videos are included in the Supplementary Information.$^\dag$

\begin{figure}[ht]
\centering
    \includegraphics[width=0.48\textwidth]{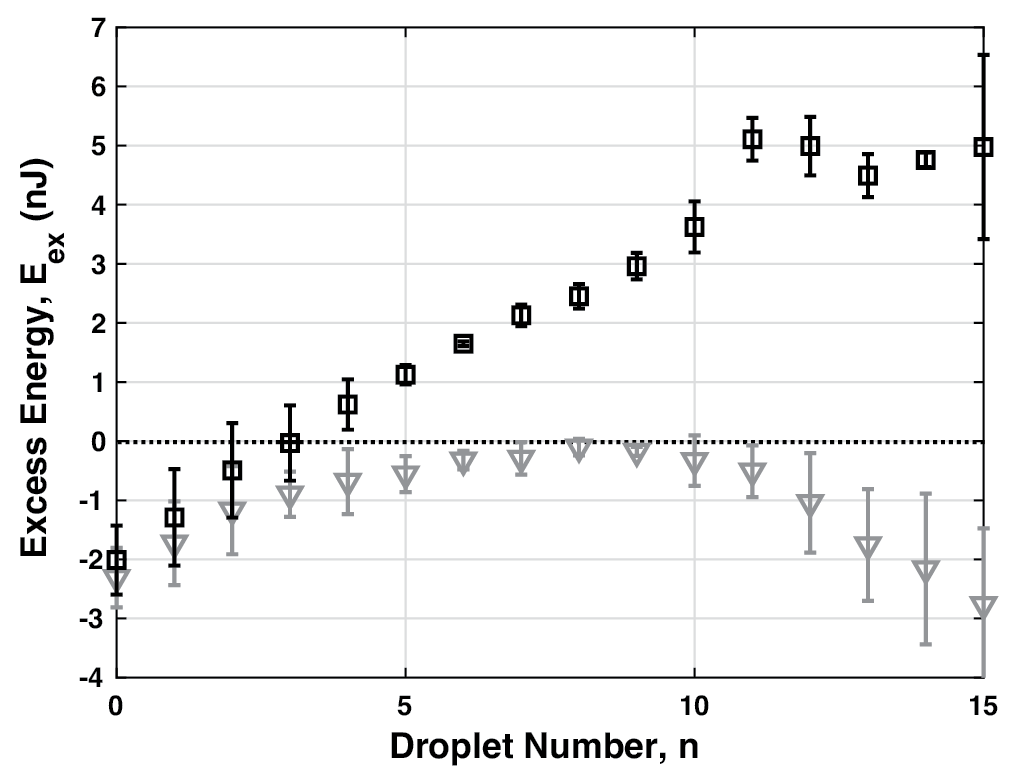}
\caption{
Droplet excess energy $E_\text{ex}$ above $U_\text{barrier}$.
$E_\text{ex}$ (black squares) and $\Delta U$ (gray triangles) averaged between two independent repeats of the same experiment.
Error bars show standard deviation between repeats.
$\Delta U < 0$ for higher droplet numbers results from early-time droplets that are already too large to stabilize.
Droplets with $E_\text{ex}>0$ are expected to escape the trapping potential, while droplets with $E_\text{ex}<0$ are expected to be able to stabilize and stop the leak.
}
\label{fig:energy_comparison}
\end{figure}

We plot the excess total energy $E_\text{ex}$ above the trapping barrier versus droplet number $n$ in Fig. \ref{fig:energy_comparison} (back squares), 
plotting the average and standard deviation between two independent repeats with the same experimental conditions. 
For comparison, we also plot the difference in potential energy from the barrier height, $\Delta U = U_\text{barrier} - U(V,P_\text{drive},a_i)$ (gray triangles), which roughly follows the shape of the potential energy barrier.

While an examination of the potential energy alone would suggest that any of the last 9-10 droplets could have stopped the leak, 
our total energetic theory makes clear that most of these droplets were growing too quickly and thus had an initial total energy too large to be caught by their respective trapping potentials.
Although the theory predicts that the leaks might have stopped 1-2 droplets earlier than observed, overall we see good agreement with the experiments, especially given the fitting approximations required to measure $m_\text{eff}$ and $dV/dt$. 
This energetic analysis clearly captures the progression toward creation of a stable capping droplet.
Importantly, this shows that the kinetic energy has a significant impact on a droplet's excess energy, contributing over half of the change in total energy from droplet to droplet as the system approaches spontaneous arrest from the dripping regime.

\section{Characterizing Spontaneous Arrest over a Broad Parameter Space}

Thus far, we have focused on a specific set of experimental conditions that reliably demonstrated an extended dripping flow regime between rivulet flow and spontaneous arrest. 
Such leaks allow us to analyze the approach to leak-stop in detail as we measure each droplet sequentially.
However, this energetic analysis can be applied more broadly across the parameter space of contact angle and hole size, including for leaks that stop directly from rivulet breakup.
(See flow sequence summary Fig. S3$^\dag$.)
In such cases, we observe that rivulet breakup immediately generates the final capping droplet, which often starts at larger volume than its equilibrium value.
In this case, the capping droplet then shrinks back into the hole and ultimately equilibrates via damped harmonic oscillations just as in the experiments analyzed above.

\begin{figure}[htb]
    \centering
    \includegraphics[width=0.48\textwidth]{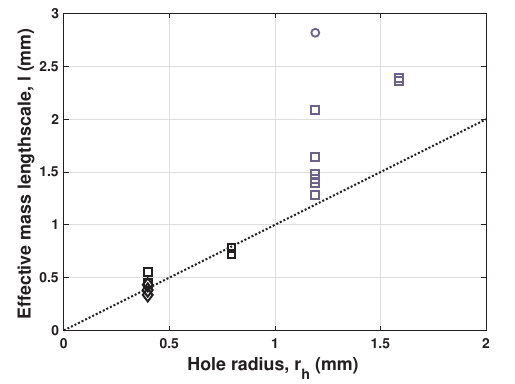}
    \caption{Effective mass length scale $l = \rho/m_\text{eff}$ of the final capping droplet versus hole radius $r_h$.
    Contact angle indicated by symbol: $\theta=149.0^\circ$ (diamonds), $\theta=108.9^\circ$ (squares), $\theta=102.1^\circ$ (circle).
    Experiments with hole sizes $r_h \leq 0.8$ mm all spontaneously arrested with total energy within 10 nJ of $U_\text{barrier}$ (black symbols), while leaks from larger hole sizes ($r_h \geq 1.2$ mm) all stopped with $E_{ex} < -25$ nJ, in some cases hundreds of nJ below $U_\text{barrier}$ (blue symbols).
    Dotted line is $l=r_h$ (slope 1).
    }
    \label{fig:effective_mass}
\end{figure}

We apply our energetic analysis to the final capping droplet in all of our experiments for which we can fit for the geometric parameters required to extract $m_{eff}$ and $dV/dt$.
Our side-view imaging setup limits us to analyzing capping droplets that are axisymmetric;
future work implementing simultaneous imaging from multiple angles to capture more complex 3D geometries will enable application of this analysis to an even broader parameter space.
We summarize the results of these energetic analyses by plotting the measured effective mass length scale $l=\rho/m_\text{eff}$ versus hole radius $r_h$ in Fig. \ref{fig:effective_mass}.
Here, plot symbols indicates the exterior tube contact angle $\theta$ and we have colored the data in two groups based on whether the leaks spontaneously arrested with an excess energy $E_{ex}$ within 10 nJ of the potential energy barrier (black symbols) or at total energies significantly below the barrier, with $E_{ex} < -25$ nJ (blue symbols).

As with the specific experiment analyzed in detail above, the effective mass length scale is approximately equal to the hole radius for smaller hole sizes, and increases with increasing hole size. 
At larger hole sizes, we observe more spread in the data, and that the length scale begins to grow more quickly than the hole radius.
This may result from more complex oscillations of the capping droplets beyond the simple fundamental oscillation mode used in our analysis.
For these larger holes, we sometimes observe higher order oscillation modes, which may complicate the effective mass measurement because the capping droplet is no longer equivalent to a simple mass-and-spring harmonic oscillator.
Expanded 3D imaging would be required to examine this further, as noted above.

For the two smallest hole sizes measured, $r_h = 0.4$ and 0.8 mm, all leaks stopped very close to the potential energy barrier, and mostly had $E_{ex}$ values clustered within a couple of nJ below zero.
These include both leaks that stopped from an extended dripping regime (all points at $r_h = 0.4$ mm) as well as two that stopped directly from rivulet breakup without a dripping regime (all points at $r_h = 0.8$ mm) (see also supplementary Fig. S3$^\dag$).
Interestingly, leaks with larger hole sizes ($r_h \geq 1.2$ mm) were able to continue leaking to much lower total energies, in some cases hundreds of nJ below the trapping potential barrier.
We observe that all of these leaks stopped directly from rivulet flow, without any dripping regime, and did not generate a self-arresting capping droplet until the system was already deeply trapped in the potential energy well. 

These results emphasize that a key prerequisite for spontaneous arrest of a leaking flow is breakup of the initial continuous flow.
As long as the fluid has a continuous path out of the hole, there is no possibility of generating an attempted capping droplet that might stop the leak, and hence leaks with more stable rivulets can persist much longer than a leak that enters a discrete dripping regime.
Fluid rivulets are a familiar sight to anyone who has watched rainwater streak down a window, and they are broadly defined as a thin stream of liquid flowing in partially wetting contact with a solid surface.\cite{davis1987thermocapillary} 
They are commonly observed in both natural and industrial processes,\cite{liu2023efficient,robertson2010numerical} and are themselves of scientific interest due to their applications in small scale fluidics.\cite{stone2004engineering}
It is established that rivulets are generally stabilized by larger contact areas,\cite{diez2009breakup} consistent with our observation that larger hole sizes and more wettable tube exteriors lead to spontaneous arrest directly from rivulet flow without an extended dripping regime. 
However, a key challenge in understanding rivulet stability stems from the fact that they display a wide range of instabilities, from localized dripping to meandering instabilities to simultaneous breakup of the entire rivulet.\cite{davis1980moving,davis1987thermocapillary,huerre1990local,tobias1998convective,myers2004stability,duprat2007absolute,diez2009breakup, daerr2011general,diez2012instability,singh2017breakup}

While our focus in this paper has been on understanding the fundamental mechanism of \textit{how} a leak can stop itself, our experiments also offer useful insights into \textit{when} a leak can stop itself by establishing an experimental system that can be used to investigate rivulet stability as a function of flow parameters including contact angle, hole size, average flow rate, interface angle, and more.
As a proof of concept, we conducted preliminary experiments cataloging rivulet instabilities and flow behavior at constant average flow rates ranging from 0.1 to 10 mL/min using our filling setup.
For these experiments, we used bare polypropylene tubes ($\theta = 102.1^\circ$) in order to facilitate more stable rivulets, again oriented vertically with respect to gravity, and with hole radii ranging from 0.4 mm to 2.4 mm.
We observed several different distinct flow behaviors across this parameter space: long stable rivulets; shorter ($\sim20$ mm), stable rivulets with end dripping; steady dripping; dripping with irregular intervals between drips; elongated ``dripulets'' that streak down the tube exterior more closely resembling traveling rivulet segments than drips; and cycles where the overall leaking behavior alternates between flowing and spontaneously-arrested states.
We include a summary of these observed rivulet flow states versus average flow rate and hole radius in supplementary Fig. S5$^\dag$.

We find that higher ($\gtrsim 3$ mL/min) and lower ($\lesssim 0.7$ mL/min) show the most predictable flow behaviors, with the higher flow rates generally supporting steady rivulet flow and lower flow rates progressing through cycles of filling to $P_\text{crit}$, flow initiation via dam-burst, draining, spontaneous arrest, and subsequent re-filling.
Intermediate flow rates were much less predictable, sometimes demonstrating as many as four different flow behaviors at the same hole size and flow rate, and evolving between different flow behaviors on a time scale of minutes.
This observed switching behavior on a time scale much larger than that of the fluid instabilities could suggest that some of the different flow states may be metastable with respect to each other or that rivulets under these experimental conditions may be exquisitely sensitive to flow conditions.
Future experiments and theoretical analyses will be required to more fully understand rivulet stability throughout this parameter space.

\section*{Conclusions}

In this work, we have seen that small fluid leaks progress through a series of flow transitions from continuous flow to spontaneous arrest with decreasing driving pressure, and ultimately can stop themselves by generating a capping droplet of fluid whose surface tension, curvature, and resulting Laplace pressure block further leakage.
These capping droplets are generated by fluid breakup events, either directly from fluid surface instabilities that break up continuous flow, or from sequential breakoffs during a dripping flow regime.
While some attempted capping droplets are unable to stop the leak and instead grow unstably until falling away as the next drip, in all cases eventually a final capping droplet is formed that 
equilibrates via lightly damped harmonic volume oscillations.
In order to understand and predict which attempted capping droplet will stop the leak, we developed a total energetic theory that accounts for both potential and kinetic energy,
in which excess kinetic energy can cause droplets to escape over a trapping potential energy barrier $U_\text{barrier}$.

Our kinetic energy analysis takes advantage of a direct analogy between the equilibrating droplets and classical mass-and-spring damped harmonic oscillators to extract key parameters, including the oscillation frequency, damping constant, and effective mass $m_\text{eff}$ of the droplet. 
By focusing our analysis on experimental conditions that reliably produce an extended dripping regime as the system approached spontaneous arrest, we characterized the progression of the leak by measuring the excess energy $E_{ex}$ of each subsequent droplet above the trapping potential.
This analysis explained why earlier droplets were not able to stop the leak despite starting with a higher Laplace pressure than the driving pressure.
Importantly, we found that the kinetic energy has a significant impact on a droplet's excess energy, contributing over half of the change in total energy per droplet as the system approaches spontaneous arrest.

These results reveal the mechanism of spontaneous arrest of a small fluid leak via generation and equilibration of a capping droplet, and our total energetic theory works well to set an upper total energy bound for a capping droplet to be able to stabilize against further leakage over a broad range of experimental parameters, including for leaks that stop directly from rivulet breakup.
Additionally, we observed that a key prerequisite for spontaneous arrest is the transition from continuous to discrete, dripping flow via jet or rivulet breakup, which can lead to leaks being dynamically stable by preventing formation of a capping droplet. 
We reported preliminary experiments examining rivulet instabilities as a function of flow rate.
These revealed a complex space of stable and metastable flow states even in our simple experimental system which will inform future experimental and theoretical studies.

Overall, our experiments highlight the rich array of physical phenomena present in the flow transitions of leaking fluids.
In addition to observing spontaneous arrest of small fluid leaks and providing some insights into rivulet instabilities, our experiments also showed regimes of irregular dripping, reminiscent of other dripping systems that have been shown to exhibit chaos. \cite{coullet2005hydrodynamical, sartorelli1994crisis, dreyer1991route}
The existence of flow start and stop cycles at low average flow rates suggests that an analysis of such fluid flows from the perspective of dynamical systems may prove fruitful.
This also suggests that one may be able to take advantage of the inherent hysteresis between flow initiation at $P_\text{crit}$ and flow stop at $P_\text{equil}$ to engineer valves with no moving parts that could be set to relieve pressure or deliver precise amounts of a fluid, for example.
Finally, we note that Eq. \ref{eqn:energy} for the potential energy of a droplet quantitatively resembles the free energy of a nucleus in classical nucleation theory.
It may be interesting in the future to explore leaking flow behaviors from the perspective of phase transformations.

\section*{Author contributions}
\textbf{CDT:} Conceptualization, Methodology, Investigation; 
\textbf{HEK:} Conceptualization, Methodology, Investigation; 
\textbf{RBT:} Investigation; 
\textbf{JMF:} Conceptualization, Methodology, Investigation; 
\textbf{KEJ:} Conceptualization, Methodology, Software, Writing - Original Draft, Supervision, Project administration, Funding acquisition;
\textbf{All authors:} Formal analysis, Writing - Review \& Editing




\section*{Acknowledgements}
We thank Chuck Extrand, Roderick V. Jensen, and the Soft and Living Materials Group at ETH Z\"{u}rich for helpful discussions, Hyeongjin Kim for assistance with preliminary imaging experiments, and Luke Moorhead for asking the practical questions that inspired the start of this work.
Acknowledgment is made to the Donors of the American Chemical Society Petroleum Research Fund for partial support of this research.
This paper is based upon work supported by the National Science Foundation under Grant No. DMR-2340259.
CDT was supported by an Edward N. Perry 1968 and Cynthia W. Wood Summer Science Research Fellowship.
RBT was supported by an Allison Davis Research Fellowship.
KEJ received research sabbatical support from both the ETH Z\"{u}rich Department of Materials and the Williams Class of 1945 World Travel Fellowship.
We acknowledge funding from the Williams College Science Center Summer Science Research program and Williams College startup funding.

\balance


\providecommand{\noopsort}[1]{}\providecommand{\singleletter}[1]{#1}%

\end{document}